\documentclass[conference]{IEEEtran}
\IEEEoverridecommandlockouts
\pdfoutput=1

\usepackage[braket, qm]{qcircuit}
\usepackage{amsmath}
\usepackage{tikz}

\newcommand*\qcontrolcolor[2]{\push{\footnotesize \tikz[baseline=(char.base)]{
                              \node[shape=circle,draw,inner sep=0.6pt,color=#1] (char) {#2};}}
                              \qw}  

\newcommand*\onecontrol{\qcontrolcolor{black}{1}}
\newcommand*\twocontrol{\qcontrolcolor{black}{2}}
\newcommand*\threecontrol{\qcontrolcolor{black}{3}}

\usepackage{cite}
\usepackage{amsmath,amssymb,amsfonts} 
\usepackage{algorithmic}
\usepackage{graphicx}
\usepackage{textcomp}
\usepackage{xcolor}

\usepackage[normalem]{ulem}
\usepackage{amssymb}
\usepackage{amsmath}
\usepackage[]{algorithm2e}
\usepackage{multirow}
\usepackage{graphicx}
\usepackage{mathtools}
\usepackage{listings}
\usepackage{booktabs}
\usepackage{xspace}
\usepackage{physics}
\usepackage{qcircuit}
\usepackage{fancyhdr}

\usepackage{rotating}
\usepackage{afterpage}
\usepackage{atbegshi}
\usepackage{floatpag}
\usepackage{zref-abspage,zref-user}
\usepackage{hyperref}

\def\BibTeX{{\rm B\kern-.05em{\sc i\kern-.025em b}\kern-.08em
    T\kern-.1667em\lower.7ex\hbox{E}\kern-.125emX}}
\begin{document}

\title{Efficient Quantum Circuit Decompositions\\via Intermediate Qudits\\
\thanks{This work is funded in part by EPiQC, an NSF Expedition in Computing, under grants CCF-1730449/1832377; in part by STAQ under
grant NSF Phy-1818914; and in part by DOE grants {DE-SC}0020289
and {DE-SC}0020331.}
}

\fancypagestyle{firststyle}{%
  \fancyhf{}%
  \renewcommand{\headrulewidth}{0pt}
  \fancyfoot[C]{\scriptsize\vspace{-2em}%
    \copyright 2020 IEEE.  Personal use of this material is permitted.  Permission from IEEE must be obtained for all other uses, in any current or future media, including reprinting/republishing this material for advertising or promotional purposes, creating new collective works, for resale or redistribution to servers or lists, or reuse of any copyrighted component of this work in other works.%
  }%
}
\pagestyle{fancy}
\renewcommand{\headrulewidth}{0pt}
\fancyhf{}
\cfoot{\thepage}

\newcommand{\TODO}[1][TODO]{{\colorbox{blue}{\textbf{\textcolor{white}{#1}}}}}
\newcommand{\hide}[1]{}

\makeatletter
\newcommand{\rotateFigurePageForLabel}[1]{%
  \zlabel{#1}%
  \AtBeginShipout{%
  \ifnum\c@page=\zref@extractdefault{#1}{abspage}{0}
    \pdfpageattr{/Rotate 90}
  \fi}
}
\makeatother

\author{\IEEEauthorblockN{Jonathan M. Baker}
\IEEEauthorblockA{\textit{Department of Computer Science} \\
\textit{University of Chicago}\\
jmbaker@uchicago.edu}
\and
\IEEEauthorblockN{Casey Duckering}
\IEEEauthorblockA{\textit{Department of Computer Science} \\
\textit{University of Chicago}\\
cduck@uchicago.edu}
\and
\IEEEauthorblockN{Frederic T. Chong}
\IEEEauthorblockA{\textit{Department of Computer Science} \\
\textit{University of Chicago}\\
chong@cs.uchicago.edu}}

\maketitle
\thispagestyle{firststyle}

\begin{abstract}
Many quantum algorithms make use of ancilla, additional qubits used to store temporary information during computation, to reduce the total execution time. Quantum computers will be resource-constrained for years to come so reducing ancilla requirements is crucial. In this work, we give a method to \textit{generate} ancilla out of idle qubits by placing some in higher-value states, called qudits. We show how to take a circuit with many $O(n)$ ancilla and design an ancilla-free circuit with the same asymptotic depth. Using this, we give a circuit construction for an in-place adder and a constant adder both with $O(\log n)$ depth using temporary qudits and no ancilla.
\end{abstract}


\section{Introduction}
Many quantum algorithms make use of ancilla, additional free bits used to store temporary information during computation which are typically returned to their original state after use. Ancilla have a variety of use cases such as to reduce the total execution time. In some cases, they can provide asymptotic improvements to the depth of circuit decompositions. This highlights an important space-time tradeoff in quantum programs - we spend extra space in the form of ancilla in order to reduce the depth of an input circuit.

Real quantum machines will have a limited number of qubits so it is important that we make the most of them to enable computation of larger, more useful problems sooner. Recently, \cite{gokhale2019asymptotic} demonstrated higher dimensional qudits could be used as a replacement for ancilla in certain circuit components to great effect. While quantum circuits are often written in terms of binary logic gates on qubits, in many quantum technologies this two-level abstraction is superficial. Superconducting qubits  \cite{low2019practical} and trapped ions \cite{liu2017transferring} have an infinite spectrum of possible states and the higher states are typically suppressed. Unfortunately, by accessing these states, the computation is subject to a greater variety of errors, in fact the number of error types scale quadratically in the computing radix \cite{gokhale2019asymptotic}. However, if qudit states are used properly, the amount gained outweighs this cost. Specifically, we use qudit states temporarily during computation while maintaining binary inputs and outputs of a circuit.

Here we propose ancilla, specifically clean ancilla, be \textit{generated} local during the decomposition of an algorithm into a quantum circuit. That is, we propose a new circuit which performs qubit-qudit compression storing the information of many qubits as a small number of qudits at the cost of some gate overhead. These compression circuits produce clean ancilla in the $\ket{0}$ state. The stored data can be retrieved later when needed since all quantum operations are reversible. Essentially, when certain groups of qubits will be unused for a long period of time, we can repurpose them by compressing them and using the produced ancilla.  This ``compression'' is a rearrangement of the stored binary values into higher states.  This allows us to store more information into the same number of physical quantum devices and free up qubits for computation.

In this work we present an application of this technique to give logarithmic depth decompositions of quantum arithmetic circuits - a carry lookahead adder and by extension addition by a constant. In Section \ref{sec:compression} we present two compression circuits for qubit-qutrit and qubit-ququart compression and evaluate advantages of various compression schemes. In Section \ref{sec:decompositions} we present our decomposition of the zero-ancilla, in-place $A+B$ adder which takes as input two registers $A$ and $B$ of qubits and possibly carryin and carryout; any fresh $\ket{0}$ states used are generated locally. We then evaluate the costs of this decomposition. Finally, we discuss various extensions to our arithmetic decomposition in Sections \ref{sec:carryin_carryout} and \ref{sec:plus_k}.
\section{Background}
\label{sec:background}

In this section we will briefly introduce the basics of quantum computing on qubits, two-level quantum systems, and then present a more general approach on qudits, $d$-level quantum systems. For a more complete introduction to quantum computing we refer the reader to \cite{MikeIke} and, for a good introduction to ternary quantum gates, \cite{di2011elementary}; our notation is a natural extension to the gates used there as well. 

\subsection{Binary Quantum Computing}
In quantum computation, we use quantum bits, or qubits, which may occupy a superposition of basis states $\ket{0}$ and $\ket{1}$. Single qubits are manipulated by applying quantum gates, such as $X$, $H$ or $Z$ which transform the state in a reversible way, unlike in typical classical computing where most operations, such as AND and OR are irreversible. In order to interact multiple qubits, such as to entangle states, gates such as the CNOT are applied. The CNOT is a controlled $X$, or NOT, operation where it is applied to a target qubit only on states where the control is in the $\ket{1}$ state.

Quantum circuits consist of a sequence of operations, also called gates, applied to a set of input qubits. The depth of a circuit is given as the length of the longest critical path from input to output and the width of a circuit is the number of qubits operated on. These circuits do not have fan-in or fan-out and so when represented each line in the circuit diagram corresponds to a single qubit and time flows from left to right from inputs to outputs.

\subsection{Multi-valued Quantum Computing}
In many quantum systems, such as the ion trap or superconducting computers, there is an infinite spectrum of discrete energy levels.  The standard binary abstraction is an artificial simplification using the first two states. Instead, we may consider qudits, $d$-dimensional qubits, to exist in a superposition of any number of these states $\ket{0}$, $\ket{1}$, $\ket{2}$, etc. We express this superposition as
$$\ket{\psi} = \alpha_0\ket{0} + \alpha_1\ket{1} + ... + \alpha_{d-1}\ket{d-1} = \sum_{i=1}^{d-1}\alpha_i\ket{i}$$
where the $\alpha_i$ are the complex amplitudes and $\abs{\alpha_i}^2$ corresponds to the probability of a qudit being measured in the $i$-th state. For an $n$ qudit system we have $d^n$ many basis states. In theory, we have access to any finite number of these levels. However, for various physical reasons, it is not often practical to use large numbers of these levels.
For example in superconducting qubits, higher energy states decay more quickly.  Also, the energy gap between states is reduced for higher states, making it harder to distinguish neighboring states and reducing their reliability.
For a complete guide to superconducting qubits we refer to \cite{krantz2019quantum}.

Qu\textit{dits} are manipulated in a similar manner to qubits, however there are many additional gates which can be used depending on $d$. In quantum circuits we inherit all of the classical reversible operations. For example, for qubits, we saw the $X$ gate was equivalent to the classical reversible NOT. For a single qudit we have access to every permutation of the $d$ basis states, or $d! - 1$ nontrivial operations. In practice, many of these operations are unnecessary and only a small number are needed to for universal computation. We make use of the increment permutations, denoted $X_{+k}$ where $+$ is addition modulo $d$, which rotates a state $\ket{i}$ to $\ket{i + k \mod d}$ and the flip permutations denoted $X_{ij}$ which flip the states $\ket{i}$ and $\ket{j}$ and leaves all others unchanged. $X_{01}$ is equivalent to the qubit $X$ gate.

Each of these operations can be extended to two qudits as a controlled operation that applies the single-qudit operation conditioned on the control qudit being in a certain state. For example, consider applying an $X_{+2}$ operation on a $d=4$ level system conditioned on a control qudit being in the $\ket{3}$ state. These controlled qudits have been physically realized and they are universal for qudit computation \cite{MS-gates}. This can be extended to any number of controls but only two-qudit gates can be directly executed on typical quantum hardware; any use of a multi-controlled gate has a decomposition into one and two qudit gates since these gates are universal. We only require a single 2-controlled gate (Toffoli-like) and its decomposition is given in \cite{di2011elementary}. We represent these gates in circuit diagrams with the control types indicated by circles with the control values inside. The applied gates, specifically the increment ($X_{+i}$) and flip gates ($X_{ij}$) will be given as a square with the type of gate inside.
\section{Qubit-Qudit Compression}
\label{sec:compression}

\begin{figure*}[t]
    \centering
    \[\input{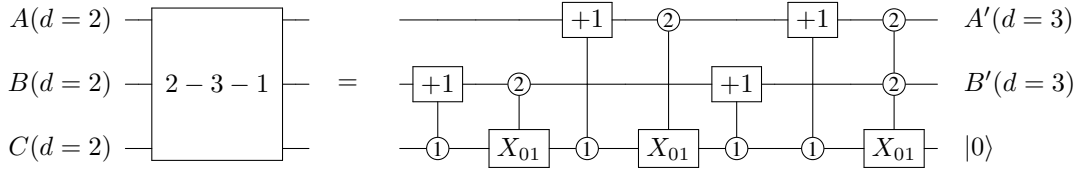}\]
    \caption{The compression of 3 qubits into 2 qutrits and an ancilla, $\ket{0}$. All $+1$ gates are done modulo 3. Using a sequence of qutrit gates, we can transform three input qubits into the desired ancilla. When A, B and C are not going to be used for a long time in the circuit, they can be temporarily repurposed as an ancilla bit elsewhere in the circuit. When we want to operate on these stored bits, we run the inverse of this circuit using \textit{any} ancilla for the third qubit.}
    \label{fig:23compression}
\end{figure*}

Typically, when using a higher radix computing paradigm, we express a circuit entirely in the specified base, that is all inputs and outputs are in the designated radix. An alternative approach is to fix the input and output radix but allow the use of higher level states temporarily during the computation, i.e. we are permitted to occupy any level up to a specified $d$ during a computation with the guarantee that we return to the specified radix.

What does this gain for us? It is known that by simply fully encoding a computation into a higher radix we obtain a constant space and time advantage over binary-only circuits. However, recently it was shown that use of these higher states can act as temporarily storage, similar to the use of an ancilla, and can convey an asymptotic reduction in circuit depth \cite{gokhale2019asymptotic}. This circuit construction, as well as other work, suggests we can obtain better circuits while using fewer qubits by accessing higher states temporarily.

We take this a step further and \textit{generate} ancilla temporarily out of input qubits in order to take advantage of previously known efficient binary circuit decompositions like that of \cite{draper2006logarithmic}. Using this method, we can reduce the number of external ancilla needed from $O(n)$ to $0$ while keeping the same asymptotic circuit depth. To do this, we allow subsets of qubits to temporarily store higher values, becoming qudits, to store the information of many qubits within a few qudits. As a concrete example, consider three qubits. There are $2^3 = 8$ total basis states while for two qutrits there are $3^2 = 9$ basis states. Therefore all the information of 3 qubits can be stored in two qutrits and the third qubit can be left in a chosen state, $\ket{0}$, a clean ancilla. We refer to this process as \textit{compression}, that is storing the information of many qu\textit{bits} in a smaller number of qu\textit{dits}.

We consider various reversible compression schemes labeled x-y-z compression, where $x$ is the radix of the input qudits, $y$ is the radix of the output qudits, and $z$ is the number of ancilla generated. Such operations exist if $x^m \le y^n$ with $0 < n < m$ and $m - n = z$ for some integers $m, n$, the number of input qudits and the number of non-ancilla outputs, respectively. Put more simply, these proposed compression circuits exist if the number of basis states of the inputs is fewer than the number of basis states of the non-ancilla outputs and the number of non-ancilla outputs is strictly smaller than the number of inputs. Ideally, we choose compression schemes with a good compression ratio, i.e. those for which $x^m / y^n \approx 1$. 

In this paper, we consider 2-3-1 and 2-4-1 compression as methods of generating ancilla for simplicity.  Many other schemes such as 2-8-2 and 3-9-1 are possible but require increasingly complex compression circuits.

\begin{table}[t]
    \centering
    \bgroup
    \setlength\tabcolsep{1.5em}
    \begin{tabular}{ccc|ccc}
        A & B & C & A' & B' & C' \\\hline
        0 & 0 & 0 & 0  & 0  & 0 \\
        0 & 0 & 1 & 2  & 2  & 0 \\
        0 & 1 & 0 & 0  & 1  & 0 \\
        0 & 1 & 1 & 0  & 2  & 0 \\
        1 & 0 & 0 & 1  & 0  & 0 \\
        1 & 0 & 1 & 2  & 1  & 0 \\
        1 & 1 & 0 & 1  & 1  & 0 \\
        1 & 1 & 1 & 1  & 2  & 0 \\
    \end{tabular}
    \egroup
    \vspace{1.5em}
    \caption{Truth table for 2-3-1 Compression}
    \label{tab:231_compression_table}
\end{table}

\subsection{2-3-1 Compression}
\label{sec:231compression}

In 2-3-1 compression we take as input three qubits and output 2 qutrits and a single ancilla, a qubit guaranteed to be in the $\ket{0}$ state. First, consider the truth table of Table \ref{tab:231_compression_table}. We note the partial function represented by this truth table is invertible, implying there exists a reversible circuit that realizes it. The third output, C', is guaranteed to be in the $\ket{0}$ state, an ancilla. By storing qubit information used infrequently we can generate an extremely useful ancilla to be used elsewhere in the circuit.

Because we ensure all inputs are binary, we do not need to consider the inputs with value $2$ to the ternary circuit. An example circuit realizing this truth table is given in Figure \ref{fig:23compression}. When a compression circuit of this type is applied, we need to keep track of which pair of qutrits encodes the three qubits, in order.
When the compressed data is needed, we can decompress by applying the inverse of this function. The inverse circuit is simply the gates in reverse order with $+1$ replaced with $-1$. Notably, this inversion requires an ancilla as input.
To retrieve the information, the inverse should be applied taking in any free ancilla and then the stored bits can be computed on as normal.

This circuit, while accomplishing what is desired, can be rather inefficient. For example, in architectures with limited connectivity this circuit requires some number of expensive communication operations since every input qubit must be adjacent at some point. Furthermore, this circuit requires the use of a two-controlled qutrit gate which is typically decomposed into a sequence of 6 two-qutrit gates and 10 single-qutrit gates \cite{di2011elementary}. In total this compression requires 22 gates, 12 two-qutrit and 10 single-qutrit gates.

\subsection{2-4-1 Compression}
\label{sec:241compression}

\begin{figure}[h]
    \centering
    \[\input{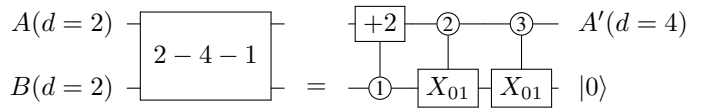}\]
    \caption{The compression of 2 qubits into a single ququart and generating an ancilla, $\ket{0}$. The $+2$ gate here is done modulo 4. This operation takes as input two qubits, A and B, and produces a single ququart and an ancilla $\ket{0}$. To do this, we need only 3 two-ququart gates. Similarly, to retrieve the stored information, we can do the inverse of this operation using \textit{any} ancilla for the second qubit.}
    \label{fig:241_compression}
\end{figure}

While 2-3-1 compression required a fairly substantial number of gates, the 2-4-1 compression circuit can convert qubit inputs into ancilla more simply and with few gates. This does not come for free. In quantum computing,  we subject our computation to a greater probability of error by using higher radix gates and by persisting for longer durations in higher energy states. In Table \ref{tab:241_compression_table}, we show that two qubits can be compressed into a single ququart and one ancilla.  2-4-1 compression is simpler than 2-3-1 compression because $2^2$ states fit evenly in a single ququart with $4$ states.  In Figure \ref{fig:241_compression}, we show a compression circuit using only 3 two-ququart gates in total, a substantial reduction over the 2-3-1 counterpart.  In the next section, we show how compression and decompression can be used to design efficient circuits requiring no ancilla.

\begin{table}[h]
    \vspace{1.5em}  
    \centering
    \bgroup
    \setlength\tabcolsep{1.5em}
    \begin{tabular}{cc|cc}
        A & B & A' & B' \\\hline
        0 & 0 & 0  & 0  \\
        0 & 1 & 2  & 0  \\
        1 & 0 & 1  & 0  \\
        1 & 1 & 3  & 0  \\
    \end{tabular}
    \egroup
    \vspace{1.5em}
    \caption{Truth table for 2-4-1 Compression}
    \label{tab:241_compression_table}
    \vspace{-1.5em}
\end{table}

\section{A+B Adder}
\label{sec:decompositions}

We now present our $A+B$ adder. This circuit takes as input two equal-sized registers of qubits, $A$ and $B$, and optionally carry-in or carry-out bits. This decomposition uses no ancilla and instead generates ancilla locally when needed by sub-components. In prior work, to achieve a logarithmic depth decomposition, $O(n)$ many ancilla were required where $n$ is the size of the input register. We will demonstrate how this efficient decomposition can be used along with our new compression technique to obtain an $O(\log n)$ depth decomposition of the same adder in-place without the extra use of ancilla.

\begin{figure}
    \centering
    \[
    \input{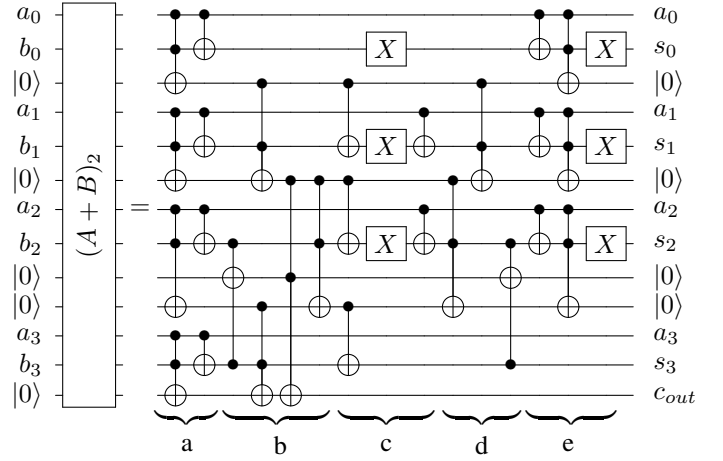}
    \]
    \caption{An adder circuit from \cite{draper2006logarithmic} on two four-bit registers $A$ and $B$ with a carry-out bit using ancilla. The sum $S$ is computed in-place on register $B$ while $A$ is untouched and the ancilla are restored to $\ket{0}$. We use this as a sub-component of our general decomposition. Each of the ancilla in this circuit can be generated from other input qubits not shown here via our compression circuits. Part a of the circuit computes carry generate and propagate for each bit position. Part b computes the carry-in for every bit position. Part c does the addition, storing the output in register $B$. Parts d and e uncompute b and a respectively, restoring the ancilla back to $\ket{0}$.}
    \label{fig:a_plus_b_prior}
\end{figure}

We first briefly review the work of \cite{draper2006logarithmic} which gives a qubit-only in-place adder with ancilla which we will refer to as $(A + B)_2$. We give the decomposition for registers of size $4$ in Figure \ref{fig:a_plus_b_prior}. One of the key contributions of this prior work is to demonstrate how, in logarithmic depth, the carry bits could be computed and used (and subsequently uncomputed to restore input ancilla back to the $\ket{0}$ state). This decomposition requires $2m - w(m) - \lfloor\log{m}\rfloor$ ancilla, where $w(m)$ is the number of $1$'s appearing in the binary expansion of the number of inputs, $m$. We will use this number later to determine how many ancilla to generate via compression. This same prior work demonstrates several variants of this circuit.  We require those with either a carry-in bit, a carry-out bit, or both.

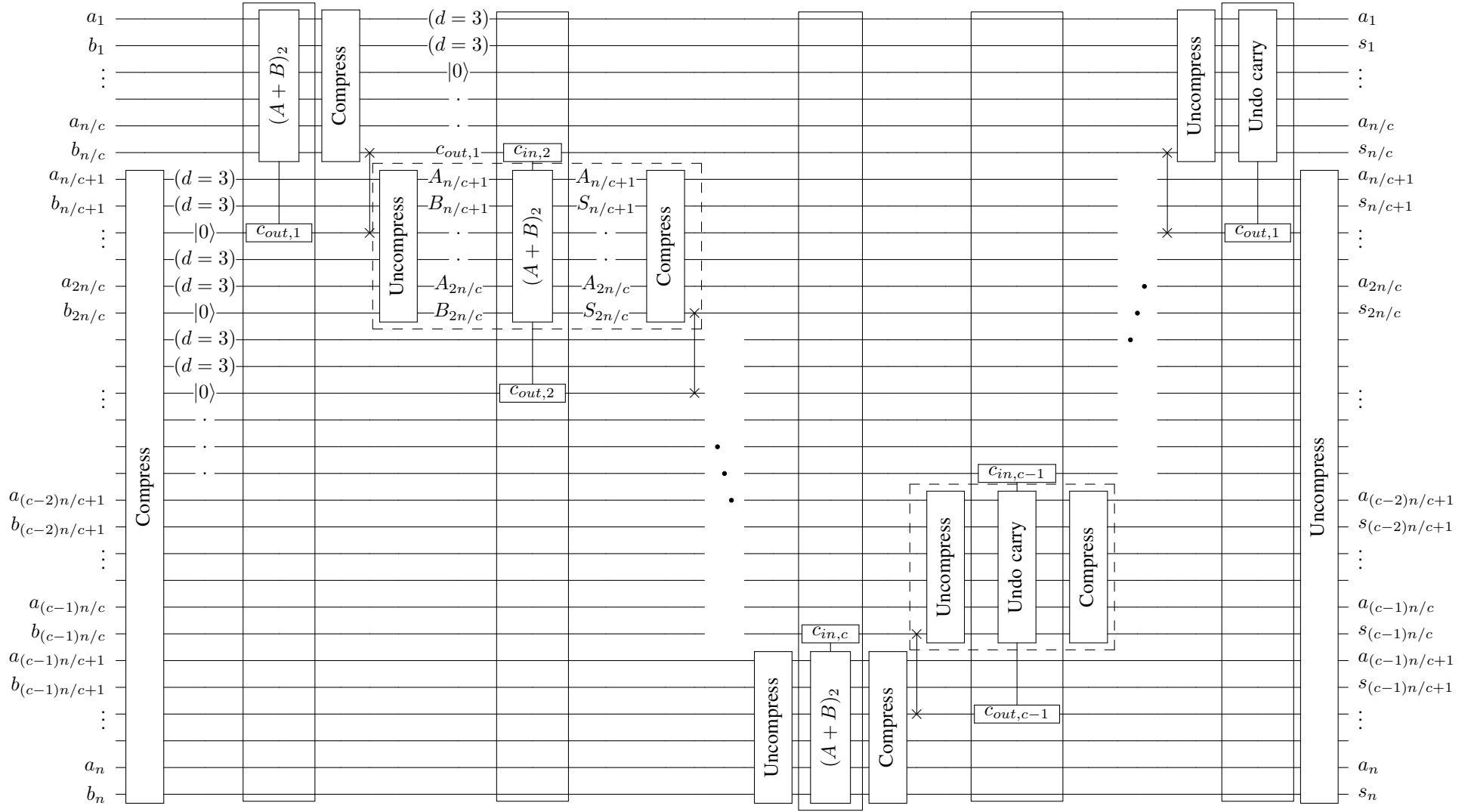
\begin{sidewaysfigure*}[p!]
    \centering
    \[%
    \newcommand*\comp{{\mbox{\rotatebox[origin=c]{90}{Compress}}}}
\newcommand*\unc{{\mbox{\rotatebox[origin=c]{90}{Uncompress}}}}
\newcommand*\abgate{{\mbox{\rotatebox[origin=c]{90}{$(A+B)_2$}}}}
\newcommand*\uac{{\mbox{\rotatebox[origin=c]{90}{Undo carry}}}}
\newcommand*\bigdot{{\tikz{\filldraw (0,0) circle (0.1em);}}}
\Qcircuit @R=0.3em @C=0.5em @!R {
\lstick{a_1}	&	\qw	&	\qw	&	\multigate{5}{\abgate}	&	\multigate{5}{\comp}	&	\qw	&	\qw	&	\push{(d=3)}\qw	&	\push{{\rule{2.9em}{0em}}}	&	\qw[-2]	&	\qw	&	\qw	&	\qw	&	\push{{\rule{1em}{0em}}}	&	\qw[-2]	&	\qw			&	\push{{\rule{2.5em}{0em}}}	&	\qw[-2]	&	\qw	&	\qw	&	\push{{\rule{3.9em}{0em}}}	&	\qw[-2]	&	\qw	&	\push{{\rule{1em}{0em}}}	&	\qw[-2]	&	\qw	&	\multigate{5}{\unc}	&	\multigate{5}{\uac}	&	\qw	&	\rstick{a_1}	\qw	&		\\
\lstick{b_1}	&	\qw	&	\qw	&	\ghost{\abgate}	&	\ghost{\comp}	&	\qw	&	\qw	&	\push{(d=3)}\qw	&	\qw	&	\qw	&	\qw	&	\qw	&	\qw	&	\qw	&	\qw	&	\qw			&	\qw	&	\qw	&	\qw	&	\qw	&	\qw	&	\qw	&	\qw	&	\qw	&	\qw	&	\qw	&	\ghost{\unc}	&	\ghost{\uac}	&	\qw	&	\rstick{s_1}	\qw	&		\\
\lstick{\vdots}	&	\qw	&	\qw	&	\ghost{\abgate}	&	\ghost{\comp}	&	\qw	&	\qw	&	\push{\ket{0}}\qw	&	\qw	&	\qw	&	\qw	&	\qw	&	\qw	&	\qw	&	\qw	&	\qw			&	\qw	&	\qw	&	\qw	&	\qw	&	\qw	&	\qw	&	\qw	&	\qw	&	\qw	&	\qw	&	\ghost{\unc}	&	\ghost{\uac}	&	\qw	&	\rstick{\vdots}	\qw	&		\\
	&	\qw	&	\qw	&	\ghost{\abgate}	&	\ghost{\comp}	&	\qw	&	\qw	&	\push{~.~}\qw	&	\qw	&	\qw	&	\qw	&	\qw	&	\qw	&	\qw	&	\qw	&	\qw			&	\qw	&	\qw	&	\qw	&	\qw	&	\qw	&	\qw	&	\qw	&	\qw	&	\qw	&	\qw	&	\ghost{\unc}	&	\ghost{\uac}	&	\qw	&		\qw	&		\\
\lstick{a_{n/c}}	&	\qw	&	\qw	&	\ghost{\abgate}	&	\ghost{\comp}	&	\qw	&	\qw	&	\push{~.~}\qw	&	\qw	&	\qw	&	\qw	&	\qw	&	\qw	&	\qw	&	\qw	&	\qw			&	\qw	&	\qw	&	\qw	&	\qw	&	\qw	&	\qw	&	\qw	&	\qw	&	\qw	&	\qw	&	\ghost{\unc}	&	\ghost{\uac}	&	\qw	&	\rstick{a_{n/c}}	\qw	&		\\
\lstick{b_{n/c}}	&	\qw	&	\qw	&	\ghost{\abgate}	&	\ghost{\comp}	&	\qswap	&	\qw	&	\push{c_{out,1}}\qw	&	\multigate{0}{c_{in,2}}\qwx[1]	&	\qw	&	\qw	&	\qw	&	\qw	&	\qw	&	\qw	&	\qw			&	\qw	&	\qw	&	\qw	&	\qw	&	\qw	&	\qw	&	\qw	&	\qw	&	\qw	&	\qswap	&	\ghost{\unc}	&	\ghost{\uac}	&	\qw	&	\rstick{s_{n/c}}	\qw	&		\\
\lstick{a_{n/c+1}}	&	\multigate{23}{\comp}	&	\push{(d=3)}\qw	&	\qw	&	\qw	&	\qw	&	\multigate{5}{\unc}	&	\push{A_{n/c+1}}\qw	&	\multigate{5}{\abgate}	&	\push{A_{n/c+1}}\qw	&	\multigate{5}{\comp}	&	\qw	&	\qw	&	\qw	&	\qw	&	\qw			&	\qw	&	\qw	&	\qw	&	\qw	&	\qw	&	\qw	&	\qw	&		&		&	\qw	&	\qw	&	\qw	&	\multigate{23}{\unc}	&	\rstick{a_{n/c+1}}	\qw	&		\\
\lstick{b_{n/c+1}}	&	\ghost{\comp}	&	\push{(d=3)}\qw	&	\qw	&	\qw	&	\qw	&	\ghost{\unc}	&	\push{B_{n/c+1}}\qw	&	\ghost{\abgate}	&	\push{S_{n/c+1}}\qw	&	\ghost{\comp}	&	\qw	&	\qw	&	\qw	&	\qw	&	\qw			&	\qw	&	\qw	&	\qw	&	\qw	&	\qw	&	\qw	&	\qw	&		&		&	\qw	&	\qw	&	\qw	&	\ghost{\unc}	&	\rstick{s_{n/c+1}}	\qw	&		\\
\lstick{\vdots}	&	\ghost{\comp}	&	\push{\ket{0}}\qw	&	\multigate{0}{c_{out,1}}\qwx[-3]	&	\qw	&	\qswap\qwx[-3]	&	\ghost{\unc}	&	\push{~.~}\qw	&	\ghost{\abgate}	&	\push{~.~}\qw	&	\ghost{\comp}	&	\qw	&	\qw	&	\qw	&	\qw	&	\qw			&	\qw	&	\qw	&	\qw	&	\qw	&	\qw	&	\qw	&	\qw	&		&		&	\qswap\qwx[-3]	&	\qw	&	\multigate{0}{c_{out,1}}\qwx[-3]	&	\ghost{\unc}	&	\rstick{\vdots}	\qw	&		\\
	&	\ghost{\comp}	&	\push{(d=3)}\qw	&	\qw	&	\qw	&	\qw	&	\ghost{\unc}	&	\push{~.~}\qw	&	\ghost{\abgate}	&	\push{~.~}\qw	&	\ghost{\comp}	&	\qw	&	\qw	&	\qw	&	\qw	&	\qw			&	\qw	&	\qw	&	\qw	&	\qw	&	\qw	&	\qw	&	\qw	&		&		&	\qw	&	\qw	&	\qw	&	\ghost{\unc}	&		\qw	&		\\
\lstick{a_{2n/c}}	&	\ghost{\comp}	&	\push{(d=3)}\qw	&	\qw	&	\qw	&	\qw	&	\ghost{\unc}	&	\push{A_{2n/c}}\qw	&	\ghost{\abgate}	&	\push{A_{2n/c}}\qw	&	\ghost{\comp}	&	\qw	&	\qw	&	\qw	&	\qw	&	\qw			&	\qw	&	\qw	&	\qw	&	\qw	&	\qw	&	\qw	&	\qw	&	~~\bigdot	&		&	\qw	&	\qw	&	\qw	&	\ghost{\unc}	&	\rstick{a_{2n/c}}	\qw	&		\\
\lstick{b_{2n/c}}	&	\ghost{\comp}	&	\push{\ket{0}}\qw	&	\qw	&	\qw	&	\qw	&	\ghost{\unc}	&	\push{B_{2n/c}}\qw	&	\ghost{\abgate}	&	\push{S_{2n/c}}\qw	&	\ghost{\comp}	&	\qswap	&	\qw	&	\qw	&	\qw	&	\qw			&	\qw	&	\qw	&	\qw	&	\qw	&	\qw	&	\qw	&	\qw	&	~\bigdot~	&		&	\qw	&	\qw	&	\qw	&	\ghost{\unc}	&	\rstick{s_{2n/c}}	\qw	&		\\
	&	\ghost{\comp}	&	\push{(d=3)}\qw	&	\qw	&	\qw	&	\qw	&	\qw	&	\qw	&	\qw	&	\qw	&	\qw	&	\qw	&	\qw	&		&		&	\qw			&	\qw	&	\qw	&	\qw	&	\qw	&	\qw	&	\qw	&	\qw	&	\bigdot~~	&		&	\qw	&	\qw	&	\qw	&	\ghost{\unc}	&		\qw	&		\\
	&	\ghost{\comp}	&	\push{(d=3)}\qw	&	\qw	&	\qw	&	\qw	&	\qw	&	\qw	&	\qw	&	\qw	&	\qw	&	\qw	&	\qw	&		&		&	\qw			&	\qw	&	\qw	&	\qw	&	\qw	&	\qw	&	\qw	&	\qw	&		&		&	\qw	&	\qw	&	\qw	&	\ghost{\unc}	&		\qw	&		\\
\lstick{\vdots}	&	\ghost{\comp}	&	\push{\ket{0}}\qw	&	\qw	&	\qw	&	\qw	&	\qw	&	\qw	&	\multigate{0}{c_{out,2}}\qwx[-3]	&	\qw	&	\qw	&	\qswap\qwx[-3]	&	\qw	&		&		&	\qw			&	\qw	&	\qw	&	\qw	&	\qw	&	\qw	&	\qw	&	\qw	&		&		&	\qw	&	\qw	&	\qw	&	\ghost{\unc}	&	\rstick{\vdots}	\qw	&		\\
	&	\ghost{\comp}	&	\push{~.~}\qw	&	\qw	&	\qw	&	\qw	&	\qw	&	\qw	&	\qw	&	\qw	&	\qw	&	\qw	&	\qw	&		&		&	\qw			&	\qw	&	\qw	&	\qw	&	\qw	&	\qw	&	\qw	&	\qw	&		&		&	\qw	&	\qw	&	\qw	&	\ghost{\unc}	&		\qw	&		\\
	&	\ghost{\comp}	&	\push{~.~}\qw	&	\qw	&	\qw	&	\qw	&	\qw	&	\qw	&	\qw	&	\qw	&	\qw	&	\qw	&	\qw	&	\bigdot~~	&		&	\qw			&	\qw	&	\qw	&	\qw	&	\qw	&	\qw	&	\qw	&	\qw	&		&		&	\qw	&	\qw	&	\qw	&	\ghost{\unc}	&		\qw	&		\\
	&	\ghost{\comp}	&	\push{~.~}\qw	&	\qw	&	\qw	&	\qw	&	\qw	&	\qw	&	\qw	&	\qw	&	\qw	&	\qw	&	\qw	&	~\bigdot~	&		&	\qw			&	\qw	&	\qw	&	\qw	&	\qw	&	\multigate{0}{c_{in,c-1}}\qwx[1]	&	\qw	&	\qw	&		&		&	\qw	&	\qw	&	\qw	&	\ghost{\unc}	&		\qw	&		\\
\lstick{a_{(c-2)n/c+1}}	&	\ghost{\comp}	&	\qw	&	\qw	&	\qw	&	\qw	&	\qw	&	\qw	&	\qw	&	\qw	&	\qw	&	\qw	&	\qw	&	~~\bigdot	&		&	\qw			&	\qw	&	\qw	&	\qw	&	\multigate{5}{\unc}	&	\multigate{5}{\uac}	&	\multigate{5}{\comp}	&	\qw	&	\qw	&	\qw	&	\qw	&	\qw	&	\qw	&	\ghost{\unc}	&	\rstick{a_{(c-2)n/c+1}}	\qw	&		\\
\lstick{b_{(c-2)n/c+1}}	&	\ghost{\comp}	&	\qw	&	\qw	&	\qw	&	\qw	&	\qw	&	\qw	&	\qw	&	\qw	&	\qw	&	\qw	&	\qw	&		&		&	\qw			&	\qw	&	\qw	&	\qw	&	\ghost{\unc}	&	\ghost{\uac}	&	\ghost{\comp}	&	\qw	&	\qw	&	\qw	&	\qw	&	\qw	&	\qw	&	\ghost{\unc}	&	\rstick{s_{(c-2)n/c+1}}	\qw	&		\\
\lstick{\vdots}	&	\ghost{\comp}	&	\qw	&	\qw	&	\qw	&	\qw	&	\qw	&	\qw	&	\qw	&	\qw	&	\qw	&	\qw	&	\qw	&		&		&	\qw			&	\qw	&	\qw	&	\qw	&	\ghost{\unc}	&	\ghost{\uac}	&	\ghost{\comp}	&	\qw	&	\qw	&	\qw	&	\qw	&	\qw	&	\qw	&	\ghost{\unc}	&	\rstick{\vdots}	\qw	&		\\
	&	\ghost{\comp}	&	\qw	&	\qw	&	\qw	&	\qw	&	\qw	&	\qw	&	\qw	&	\qw	&	\qw	&	\qw	&	\qw	&		&		&	\qw			&	\qw	&	\qw	&	\qw	&	\ghost{\unc}	&	\ghost{\uac}	&	\ghost{\comp}	&	\qw	&	\qw	&	\qw	&	\qw	&	\qw	&	\qw	&	\ghost{\unc}	&		\qw	&		\\
\lstick{a_{(c-1)n/c}}	&	\ghost{\comp}	&	\qw	&	\qw	&	\qw	&	\qw	&	\qw	&	\qw	&	\qw	&	\qw	&	\qw	&	\qw	&	\qw	&		&		&	\qw			&	\qw	&	\qw	&	\qw	&	\ghost{\unc}	&	\ghost{\uac}	&	\ghost{\comp}	&	\qw	&	\qw	&	\qw	&	\qw	&	\qw	&	\qw	&	\ghost{\unc}	&	\rstick{a_{(c-1)n/c}}	\qw	&		\\
\lstick{b_{(c-1)n/c}}	&	\ghost{\comp}	&	\qw	&	\qw	&	\qw	&	\qw	&	\qw	&	\qw	&	\qw	&	\qw	&	\qw	&	\qw	&	\qw	&		&		&	\qw			&	\multigate{0}{c_{in,c}}\qwx[1]	&	\qw	&	\qswap	&	\ghost{\unc}	&	\ghost{\uac}	&	\ghost{\comp}	&	\qw	&	\qw	&	\qw	&	\qw	&	\qw	&	\qw	&	\ghost{\unc}	&	\rstick{s_{(c-1)n/c}}	\qw	&		\\
\lstick{a_{(c-1)n/c+1}}	&	\ghost{\comp}	&	\qw	&	\qw	&	\qw	&	\qw	&	\qw	&	\qw	&	\qw	&	\qw	&	\qw	&	\qw	&	\qw	&	\qw	&	\qw	&	\multigate{5}{\unc}			&	\multigate{5}{\abgate}	&	\multigate{5}{\comp}	&	\qw	&	\qw	&	\qw	&	\qw	&	\qw	&	\qw	&	\qw	&	\qw	&	\qw	&	\qw	&	\ghost{\unc}	&	\rstick{a_{(c-1)n/c+1}}	\qw	&		\\
\lstick{b_{(c-1)n/c+1}}	&	\ghost{\comp}	&	\qw	&	\qw	&	\qw	&	\qw	&	\qw	&	\qw	&	\qw	&	\qw	&	\qw	&	\qw	&	\qw	&	\qw	&	\qw	&	\ghost{\unc}			&	\ghost{\abgate}	&	\ghost{\comp}	&	\qw	&	\qw	&	\qw	&	\qw	&	\qw	&	\qw	&	\qw	&	\qw	&	\qw	&	\qw	&	\ghost{\unc}	&	\rstick{s_{(c-1)n/c+1}}	\qw	&		\\
\lstick{\vdots}	&	\ghost{\comp}	&	\qw	&	\qw	&	\qw	&	\qw	&	\qw	&	\qw	&	\qw	&	\qw	&	\qw	&	\qw	&	\qw	&	\qw	&	\qw	&	\ghost{\unc}			&	\ghost{\abgate}	&	\ghost{\comp}	&	\qswap\qwx[-3]	&	\qw	&	\multigate{0}{c_{out,c-1}}\qwx[-3]	&	\qw	&	\qw	&	\qw	&	\qw	&	\qw	&	\qw	&	\qw	&	\ghost{\unc}	&	\rstick{\vdots}	\qw	&		\\
	&	\ghost{\comp}	&	\qw	&	\qw	&	\qw	&	\qw	&	\qw	&	\qw	&	\qw	&	\qw	&	\qw	&	\qw	&	\qw	&	\qw	&	\qw	&	\ghost{\unc}			&	\ghost{\abgate}	&	\ghost{\comp}	&	\qw	&	\qw	&	\qw	&	\qw	&	\qw	&	\qw	&	\qw	&	\qw	&	\qw	&	\qw	&	\ghost{\unc}	&		\qw	&		\\
\lstick{a_{n}}	&	\ghost{\comp}	&	\qw	&	\qw	&	\qw	&	\qw	&	\qw	&	\qw	&	\qw	&	\qw	&	\qw	&	\qw	&	\qw	&	\qw	&	\qw	&	\ghost{\unc}			&	\ghost{\abgate}	&	\ghost{\comp}	&	\qw	&	\qw	&	\qw	&	\qw	&	\qw	&	\qw	&	\qw	&	\qw	&	\qw	&	\qw	&	\ghost{\unc}	&	\rstick{a_{n}}	\qw	&		\\
\lstick{b_{n}}	&	\ghost{\comp}	&	\qw	&	\push{{\rule{2.9em}{0em}}}	&	\qw[-2]	&	\qw	&	\qw	&	\qw	&	\qw	&	\qw	&	\qw	&	\qw	&	\qw	&	\qw	&	\qw	&	\ghost{\unc}			&	\ghost{\abgate}	&	\ghost{\comp}	&	\qw	&	\qw	&	\qw	&	\qw	&	\qw	&	\qw	&	\qw	&	\qw	&	\qw	&	\push{{\rule{2.9em}{0em}}}	&	\ghost{\unc}\qw[-2]	&	\rstick{s_{n}}	\qw	&	\push{{\rule{2em}{0em}}}	%
						{\gategroup{1}{4}{30}{4}{.7em}{-}}						{\gategroup{7}{7}{12}{12}{.7em}{--}}				{\gategroup{1}{9}{30}{9}{.7em}{-}}																		{\gategroup{1}{17}{30}{17}{.7em}{-}}				{\gategroup{19}{19}{24}{22}{.7em}{--}}				{\gategroup{1}{21}{30}{21}{.7em}{-}}														{\gategroup{1}{28}{30}{28}{.7em}{-}}								%
}%
    \]%
    \caption{Our $A + B$ adder that uses no external ancilla.  The variant shown here for $c=5$ uses 2-3-1 compression to generate one ancilla (marked as $\ket{0}$) for every three unused qubits, storing their values in two qutrits (marked as $d=3$).  A box is drawn around every $(A+B)_2$ and Undo carry gate to indicate that they use all the generated ancilla across the circuit.  $c_{out,i}$ or $c_{in,i}$ is included on some of the gates to indicate when the carry-in and carry-out versions are used and on which ancilla the carry-out is stored.  The SWAP gates (pairs of $\times$ in the diagram) simply move a carry-out bit to another ancilla where it is used as the next carry-in. The two blocks of gates shown with dashed lines are repeated $c-2 = 3$ times along the diagonal indicated.  If 2-4-1 compression is used, an ancilla is generated for every two unused qubits so only $c=4$ blocks are needed.  The depth of this circuit is $O(\log{n})$.}
    \label{fig:a_plus_b_blocks}
    \rotateFigurePageForLabel{aplusbblocks}
\end{sidewaysfigure*}

We will now present our decomposition shown in Figure \ref{fig:a_plus_b_blocks}. Let $A = (a_1a_2\dots a_{n})$ and $B = (b_1b_2\dots b_n)$ be the input registers with $a_1, b_1$ the least significant bits of each register. We divide these registers into $c$ blocks $R_1, \dots, R_c$ each of size $2n / c$.  We assume for clarity that $n$ is a multiple of $c$ but our constructions will work for any $n$, with one additional block containing the remaining $2(n \mod c)$ qubits. Take $R_i = (a_{(i-1)(c/n)+1}b_{(i-1)(c/n)+1}\dots a_{i(c/n)}b_{i(c/n)})$ then notice for $i > 1$ we can perform an addition circuit $(A+B)_2$ with carry-in and carry-out on block $R_i$ in $O(\log(n/c)) = O(\log n)$ depth by generating the proper number of ancilla out of the other input qubits, specifically $2(n/c) - w(n/c) - \lfloor\log{n/c}\rfloor$ ancilla. We will assume a worst case scenario of $2n/c$ ancilla to simplify the analysis. Suppose we are performing $(A + B)_2$ on block $R_i$ while every other block is unused. We can perform compression on the currently unused qubits in all other blocks $\{R_j | j \neq i\}$ to obtain generated ancilla which can then be used by the current adder subcircuit.

Recall 2-3-1 compression takes 3 qubits and outputs a single ancilla. Let a 2-3-1 \textit{Compress circuit} be a circuit which takes any number of qubits $m$ as input and applies 2-3-1 compression to triplets resulting in $\lfloor m / 3 \rfloor$ ancilla. Then applying 2-3-1 compression to all qubits in $\{R_j | j \neq i\}$ we obtain $\lfloor (c - 1)2n/3c \rfloor$ ancilla. We now have constraints on what the constant $c$ should be for our decomposition to be feasible. That is we must have $\lfloor (c-1)2n/3c\rfloor \ge 2n/c$. Because we must store intermediate carry values between each $(A+B)_2$, we will actually require an additional $c-1$ ancilla, giving us $\lfloor (c - 1)2n/3c \rfloor \ge 2n/c + c - 1$.  By solving the inequality, this implies our construction is feasible for $c = 5$ and $n \ge 30$.  An alternative adder that is ancilla-free but does not scale well asymptotically, like an $O(n)$-depth adder \cite{cuccaro2004new}, may be used where our construction is infeasible on small problem sizes with $n < 30$.

The circuit construction now goes as follows, first considering the case when we have no carry-in and no carry-out. To add in these additional features requires only minor adjustments, discussed later. First, we compress the qubits in blocks $\{R_j | j \neq 1\}$. Then we apply $(A + B)_2$ with carry-out to the block $R_1$ using the newly generated ancilla. The compression block is constant depth ($O(1)$) and the adder is logarithmic depth ($O(\log(n/c)) = O(\log n)$). The qubits $b_1, \dots, b_{n/c}$ now store the first $n/c$ bits of the addition, $s_1, \dots, s_{n/c}$.  Also note the adder circuit restores all ancilla (except the carry-out) to $\ket{0}$. Then, apply a compression block to $R_1$. Swap the carry-out, $c_{out,1}$, to any of the ancilla generated to hold on to whether a carry should be applied to the next block (these carries are where the additional $c-1$ term come from above). Next, we uncompress all of the bits in $R_2$
so we can apply $(A+B)_2$ with carry-out \textit{and} carry-in ($c_{in} = c_{out,1}$) to block $R_2$ using the other generated ancilla. We repeat this process until the last block, $R_c$. In this case, since we do not have a carry-out bit we apply $(A+B)_2$ with only carry-in ($c_{in} = c_{out, c-1}$). 

We have now computed the sum $A+B$ and now must cleanup the intermediate carry bits.  This can be done by working in reverse to uncompute each carry-out without undoing the addition.
One intuitive way would be to simply apply the inverse of the $(A+B)_2$ circuit we applied to block $R_{c-1}$ which will uncompute the addition and $c_{out, c-1}$ and then re-apply it \textit{without} carry-out. Now the ancilla storing $c_{out, c-1}$ is restored to $\ket{0}$. We repeat this process on each of the blocks in reverse order. Finally, after $c_{out,1}$ has been uncomputed and the ancilla restored to $\ket{0}$, we uncompress all of the qubits. The resulting output will be the sum $S$ in register $B$ with register $A$ left unchanged from the input.

Uncomputing the intermediate carry-out bits can be improved dramatically by noticing that by applying the inverse of $(A+B)_2$ with carry-in and carry-out and the subsequently applying $(A+B)_2$ with only carry-in is unnecessary. Instead we can uncompute the carry-out by only applying the inverse of the second half of $(A+B)_2$ with carry-out and then executing the second half of $(A+B)_2$ with a few extra gates in Figure \ref{fig:a_plus_b_prior}d to cancel the carry-out.

Earlier, we show our decomposition only works when $c = 5$ using 2-3-1 compression. However, due to page size constraints, we do not show some of the repeated blocks in Figure \ref{fig:a_plus_b_blocks}. The block of gates surrounded by a dashed line is simply repeated in a block diagonal pattern indicated by the ellipsis. If we instead used 2-4-1 compression, the factor of $3$ in the earlier inequality would be replaced with $2$ making $c=4$ feasible with a constraint of $n \geq 12$.

Our decomposition performs addition in-place with zero ancilla, taking advantage of qutrits (qudits in general) to obtain ancilla instead of extra qubits for ancilla. Each of the $(A+B)_2$ blocks has depth $O(\log n)$ for input register size $n$ and we perform only a constant $2c-1$ of them so our decomposition also has $O(\log n)$ depth.

\subsection{Carry-in and Carry-out}
\label{sec:carryin_carryout}

We can extend the above decomposition to allow for carry-in quite simply. When computing the $(A+B)_2$ and Undo carry on $R_1$ we simply use the $(A+B)_2$ circuit with carry-in. Similarly, we can allow for carry-out by simply substituting an $(A+B)_2$ with carry-in \textit{and} carry-out on block $R_c$. 

\subsection{+K Adder}
\label{sec:plus_k}
The method used to construct the $A+B$ adder shown above can be applied to any circuit that can be divided into blocks while only needing to pass a constant number of bits to the input of the following block.  One example that follows from $A+B$ is the $+K$ adder.  The $+K$ adder acts on a single register of qubits $B$ and computes the sum $B+K$ in-place where $K$ is a classical constant known when creating the circuit.

The design of our $+K$ adder will use as subcircuits the $(+K)_2$ circuit derived from $(A+B)_2$ from \cite{draper2006logarithmic} and described earlier.  The design of $(+K)_2$ is the same as $(A+B)_2$ except the qubits of register $A$ are removed and all CNOT gates with a control on $a_i$ are removed and only replaced with $X$ gates if $k_i = 1$.  Similarly, the Toffoli gates (controlled-controlled-not gates) are removed and replaced with CNOT gates in the same way.  Depending on the value of $K$, some of the ancilla may also be removed but in the worst case, $(+K)_2$ may still require $2n/c - w(n/c) - \lfloor\log{n/c}\rfloor - 1$ ancilla for input size $n/c$ which we upper bound by $2n/c$.  The circuit still has $O(\log n)$ depth.

We use the same diagonal block structure as $A+B$ but now we define $R_i = (b_{(i-1)(c/n)+1}\dots b_{i(c/n)})$.  At step $i$, the number of ancilla generated by applying 2-3-1 compression to all qubits in $\{R_j | j \neq i\}$ is $\lfloor (c - 1)n/3c \rfloor$.  From this, we obtain the inequality $\lfloor (c - 1)n/3c \rfloor \geq 2n/c + c - 1$ which determines when there are enough unused qubits to generate the required ancilla.  The extra $c - 1$ ancilla are needed to store intermediate carry values.  When we solve this inequality, we find that $c=8$ blocks are required and the circuit will only have enough ancilla when $n \geq 168$.  Both the number of blocks and the minimum $n$ are larger than for $A+B$ because the input to $+K$ is only a single register so the ancilla required per input qubit is doubled, resulting in a higher minimum $n$.

2-3-1 compression is not the only option.  If we use 2-4-1 compression instead, more ancilla can be generated per input qubit and we obtain the inequality $\lfloor (c - 1)n/2c \rfloor \geq 2n/c + c - 1$.  The solution to this tells us that the minimum $c=6$ and we can use the circuit for $n \geq 60$.

\section{Conclusion}
\label{sec:conclusion}

We have shown a new use of qudits to generate ancilla in-place and its application to the class of quantum circuits that can be split into blocks. We give a new construction for an in-place addition circuit that uses no ancilla but still obtains the same $O(\log n)$ asymptotic depth as the qubit circuit it was based on that needed $O(n)$ ancilla.  The new circuit can be used as a drop-in replacement in algorithms to use significantly fewer total qubits.  These results should encourage further work in qudit-assisted quantum computing. We are particularly interested in validating these designs on hardware as more quantum machines are built. This would allow us to better evaluate the space and reliability tradeoffs of using higher radix quantum states.
\section{Acknowledgements}
\label{sec:acknowledgements}

We thank Craig Gidney and Pranav Gokhale for helpful discussions.

\bibliographystyle{ieeetr}
\bibliography{ref}

\end{document}